# Infrared Reflection Absorption Spectroscopy Setup with Incidence Angle Selection for Surfaces of Non-Metals

David Rath,[1] Vojtěch Mikerásek,[1] Chunlei Wang,[1] Moritz Eder,[1] Michael Schmid,[1] Ulrike Diebold,[1] Gareth S. Parkinson,[1] and Jiří Pavelec[1]

[1] *TU Wien , Institute of Applied Physics, Wiedner Hauptstraße 8-10/134, 1040 Vienna, Austria*

Infrared Reflection Absorption Spectroscopy (IRAS) on dielectric single crystals is challenging because the optimal incidence angles for light-adsorbate interaction coincide with regions of low IR reflectivity. Here, we introduce an optimized IRAS setup that maximizes the signal-to-noise ratio for non-metals. This is achieved by maximizing light throughput, and by selecting optimal incidence angles that directly impact the peak heights in the spectra. The setup uses a commercial FTIR spectrometer and is usable in ultra-high vacuum (UHV). Specifically, the design features sample illumination and collection mirrors with a high numerical aperture inside the UHV system, and an adjustable aperture to select the incidence angle range on the sample. This is important for p-polarized measurements on dielectrics, because the peaks in the spectra reverse direction at the Brewster angle (band inversion). The system components are connected precisely via a single flange, ensuring long-term stability. We studied the signal-to-noise variation in p-polarized IRAS spectra for one monolayer of CO on $TiO_2(110)$ as a function of incidence angle range, where a maximum signal-to-noise ratio of 70 was achieved at 4 cm$^{-1}$ resolution in five minutes measurement time. The capabilities for s-polarization are demonstrated by measuring one monolayer $D_2O$ adsorbed on a $TiO_2(110)$ surface, where a SNR of 65 was achieved at a $\Delta R/R_0$ peak height of $1.4\times10^{-4}$ in twenty minutes.

## I. INTRODUCTION

Infrared vibrational spectroscopy is a versatile and widely used tool to identify molecular species via their characteristic vibrational frequencies. This technique is particularly useful in catalysis research, as it can identify surface-bound reaction intermediates under *operando* conditions.[1,2] While much work is performed on powder-based catalysts, their structural complexity makes it difficult to link the adsorbed species to a particular active site. For this reason, many groups utilize the so-called surface-science approach, where experiments are performed on well-defined single crystal samples in a tightly controlled ultra-high vacuum (UHV) environment.[3] In such experiments, IR light is reflected once off the sample surface, and the spectrum is obtained with a Fourier transform infrared (FTIR) spectrometer.[4,5] The method is called Infrared Reflection Absorption Spectroscopy (IRAS or IRRAS), Reflection Absorption IR Spectroscopy (RAIS or RAIRS) or External Reflection Spectroscopy (ERS). Initially, IRAS was developed for metal surfaces,[6,7] where surface selection rules dictate that only vibrations with dipole changes perpendicular to the metal surface are infrared active.[8–10] On metallic substrates, the formation of an image dipole enhances the interaction of the dipole moment with the electric field. This has a maximum at grazing incidence angles of the light,[11] and allows low coverages of adsorbates to be routinely detected. On dielectrics, however, IRAS experiments are less sensitive,[12–14] partly because a weaker surface electric field interacts with the dipole, and partly because the surface field maximum overlaps with a region of comparatively poor infrared reflectivity.[8,11] An elegant way to circumvent this issue is to grow thin films of the dielectric in question on a metal support, and significant progress has been made using this approach.[15–19] Often, this is not feasible, or the thin films are affected by the proximity of the metallic layer, both structurally and electronically, necessitating the study of bulk single crystals. Experiments performed on such samples have the advantage that dipoles parallel to the surface can be studied using s-polarized light.[20] However, low signals (typically one to two orders of magnitude lower than on metals[4,21]) necessitate long measurement times,[22] particularly for sub-monolayer adsorbate coverages.[12–14]

In this paper, we describe an IRAS setup uniquely designed for studying low coverages of adsorbates on dielectric single crystals. While most current IRAS measurement systems employ fixed grazing incidence angles of 80°–85°, several studies[23–26] demonstrated that altering the incidence angle of the IR light can enhance the IRAS peak heights and the signal-to-noise ratio (SNR). Our setup provides an adjustable incidence angle range, allowing us to maximize the peak heights obtained for each material we study. We also optimized the optical throughput, which effectively minimizes noise. Hence, our system achieves an SNR of approximately 70 within minutes of measurement time for monolayer coverage.

The paper is organized as follows: The first section describes the underlying theory of IR reflectivity and the optical design. Next, we describe the mechanical design, focusing on its integration with an existing surface chemistry UHV chamber[27] that utilizes a custom-built molecular beam[28] for temperature-programmed desorption (TPD) experiments. This molecular beam, particularly its precise beam spot, imposes design constraints for the controlled illumination of the adsorbate-covered area illumination on the sample. Finally, we evaluate our IRAS setup by measuring CO and $D_2O$ molecules adsorbed on a $TiO_2$(110) surface, thus demonstrating the system capabilities.

## II. REFLECTIVITY OF METAL OXIDES

IRAS measurements on metal single crystals are sensitive to adsorbate vibrations with the dipole moment perpendicular to the surface due to the dipole enhancement effect, a high vertical surface electric field at grazing incidence angles, and the near-unity reflectivity of both polarizations in the mid-IR regime. The adsorbate-light interaction of dipole moments oriented parallel to the surface is suppressed due to the missing horizontal surface electric field; this is also known as the surface selection rule.[8–10]

For metal oxides and other dielectrics, the dipole enhancement effect is small or negligible. On these materials, the surface electric field is small compared to metals. The adsorbate-light interaction optimum is at lower incidence angles than on metals,[8,11] where the reflectivity of the substrate is low (FIG. 1). An interaction is present for all dipole moment orientations and polarizations.

The incidence angle-dependent reflectivity $R(\theta)$ of s- and p-polarized light on a flat surface can be calculated using the Fresnel equations and the material-specific complex refractive index $\hat{n} = n + ik$, where $n$ is the real refractive index and $k$ is the extinction coefficient. In FIG. 1, we show the results of the Fresnel equations for $TiO_2$ at a wavenumber $\tilde{\nu} = 2178$ cm$^{-1}$ (this is a typical value for the CO stretch; see chapter V.B.1). It should be noted that $TiO_2$ is birefringent,[29] but in the following, this aspect is neglected in the calculations. For p-polarized light, the reflectivity is zero at the Brewster angle $\theta_B$. It should be noted that these calculations are valid for a single ray and reflection at one point on the sample.

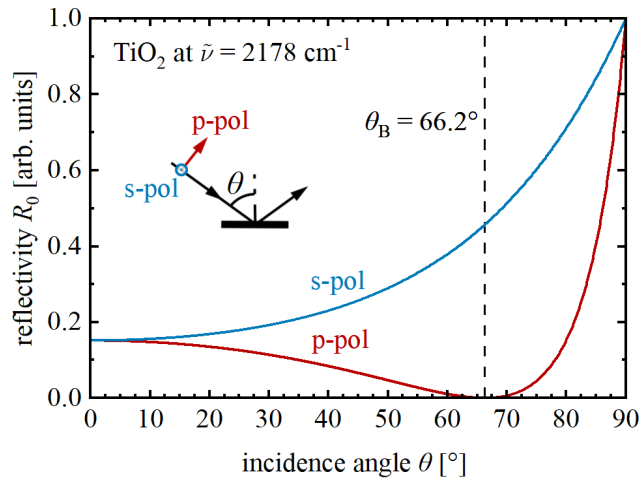

FIG. 1 Calculated angle-dependent Fresnel reflection on $TiO_2$. S-polarized (blue) and p-polarized reflection (red) at $\tilde{\nu} = 2178$ cm$^{-1}$ for a $TiO_2$ substrate with a complex refractive index[30,31] $\hat{n} = 2.27 + i\,0.002$. The dashed line indicates the Brewster angle $\theta_B$.

IRAS measures differences in the reflectivity of the adsorbate-covered surface ($R$) with respect to the pristine surface ($R_0$). In the following sections, we will refer to the normalized reflectivity difference defined as

$$\frac{\Delta R}{R_0} = \frac{R - R_0}{R_0}, \quad (1)$$

as a function of the wavenumber $\tilde{\nu}$ and the incidence angle $\theta$. Langreth[32] presented a way to calculate the incidence angle-dependent normalized reflectivity difference $\Delta R(\theta,\tilde{\nu})/R_0(\theta,\tilde{\nu})$ for an adsorbate based on the surface polarizability. Adaption of these formulas and expansion by the polarizability modeled with a Lorentzian oscillator according to equation (8) in Tobin's work,[33] forms the basis for our calculation of the normalized reflectivity difference defined in equation (1) (see equations (1) to (6) in the Supplementary Information).

For the calculations in this work, the static electronic polarizability of the adsorbate was taken as $\alpha_e = 0$, and the vibrational polarizability as $\alpha_v = 3 \times 10^{-26}$ cm³. The density of adsorbates was assumed to be $N_s = 5.2 \times 10^{14}$ cm$^{-2}$, and the line width $\gamma$ corresponds to 5 cm$^{-1}$. We consider the normalized reflectivity difference at the resonance frequency (peak height), $\tilde{\nu}_0 = \tilde{\nu} = 2178$ cm$^{-1}$. For the substrate, we assume the complex refractive index of TiO$_2$,[30,31] $\hat{n} = 2.27 + i\, 0.002$.

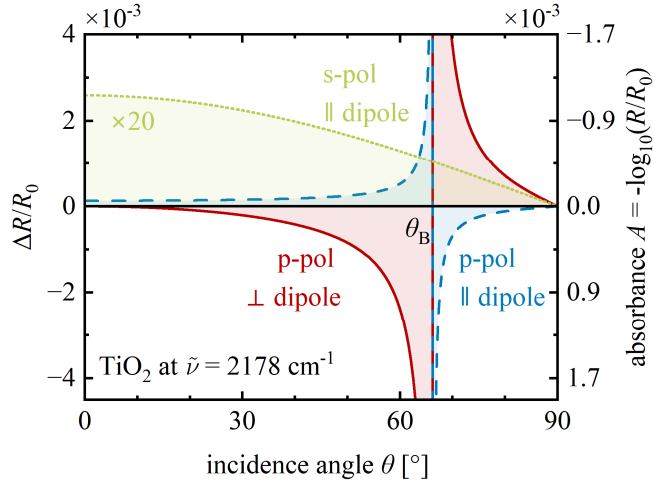

FIG. 2 Calculated normalized reflectivity difference (peak height) $\Delta R/R_0$ of an adsorbate on a TiO$_2$ surface. Three configurations of the surface adsorbate and the incoming light polarization are shown: horizontal dipole moment perpendicular to the plane of incidence and s-polarized light (green, dotted), horizontal dipole in the plane of incidence (blue, dashed), and vertical dipole moment (red); the latter two are detectable with p-polarization. Due to the low signal, the s-polarized data was multiplied by 20. $\theta_B$ denotes the Brewster angle.

FIG. 2 shows the calculated normalized reflectivity difference $\Delta R/R_0$ caused by the adsorbate as a function of the incidence angle $\theta$ under these assumptions. A dipole moment parallel to the surface but perpendicular to the incidence plane can be detected only with s-polarization. For this case (green, dotted line), $\Delta R/R_0$ remains positive because the adsorbate causes an increase in the reflected intensity at the resonance. With increasing incidence angle $\theta$, $\Delta R/R_0$ decreases. In contrast, $\Delta R/R_0$ for p-polarized light shows a $1/x$-like singularity at the Brewster angle $\theta_B$ for an ideal (non-lossy) dielectric. This is valid for both horizontally and vertically oriented dipoles in the plane of incidence. The singularity arises because $\Delta R$ has a zero of order one and $R$ a zero of order two at the Brewster angle. The change of sign at the Brewster angle $\theta_B$ is known as band inversion.[25,34–36] In the case of materials with a non-negligible imaginary part $k$ of the refractive index, the normalized reflectivity difference at the resonance frequency $\tilde{\nu}_0$ is zero close to the Brewster angle, and a Fano-like line shape is observed in the IR spectra instead of a Lorentzian peak.[37] Then, $\Delta R/R_0$ at the resonance frequency remains finite for all angles but has maxima and minima with a large magnitude near the Brewster angle (not shown in FIG. 2). For p-polarization and a perpendicular dipole moment, positive absorbance ($\Delta R/R_0 < 0$, as for adsorbates on metals) occurs only at $\theta < \theta_B$, not at grazing reflection, and $\Delta R/R_0$ approaches zero at both ends of the 0–90° range. On dielectrics, p-polarization can also detect a dipole moment oriented parallel to the surface and in the incidence plane. In this case, positive absorbance ($\Delta R/R_0 < 0$) occurs at grazing angles; $\Delta R/R_0$ again approaches zero at $\theta \to 90°$. $\Delta R/R_0$ is positive below the Brewster angle. At perpendicular incidence, the difference between s- and p-polarization becomes meaningless, thus the reflectivity for a parallel dipole moment oriented along the direction of the electric field is the same for s- (multiplied by 20 in FIG. 2) and p-polarization at $\theta = 0$. A comparison of horizontal and vertical dipole orientations for p-polarization shows almost an order of magnitude stronger signals for the vertical orientation, except for small incidence angles.

### III. DESIGN FUNDAMENTALS

Designing a highly sensitive IRAS setup relies on two fundamental objectives: minimizing noise levels and maximizing the peak heights in the spectrum. Improving both parameters results in a high SNR necessary for measuring low coverages of adsorbates. The following sections discuss our approaches to fulfill both requirements: low noise levels are primarily gained by maximizing the optical throughput, whereas increased peak heights are obtained by tailoring the light incidence angle on the sample.

#### A.   Throughput Optimization

Our IRAS setup is designed specifically to detect adsorbates with low coverages on dielectrics. To improve the SNR, one could conduct measurements with more scans, which extends the measurement time. However, further adsorption/desorption on the studied surfaces can happen during the measurement, or the background of the spectra can change, e.g., due to thermal drift. We have implemented several strategies to reduce noise while keeping the measurement time at a minimum. These strategies include optimizing the beam path and carefully selecting optical components leading to high intensities on the detector to improve the SNR. Furthermore, ensuring the mechanical stability of the system reduces drift caused by movement of the focal spot on the detector or the sample and reduces noise due to environmental vibrations.

The optical throughput of the system defines the amount of light that reaches the detector after being emitted from the source and reflected off the sample surface. The implemented optimization approach can be divided into minimizing losses on the optical components and conservation of étendue.[38,39] Étendue is a geometric quantity defined as the product of illuminated surface area $A$ and the solid angle $\Omega$ spanned by the rays in the light beam at a focus, on a surface perpendicular to the principal ray. It can be interpreted as the volume in phase space of the light beam. With a given diameter of an initially almost parallel beam, $\Omega$ is inversely proportional to the focal length. Thus, a small illuminated area $A$ (restricted by the size of the crystal or, in our case, by the adsorbate-covered area) must be accompanied by a large solid angle, which requires a short focal length for illumination. For maximum throughput, the beam path should be designed without any obstructions. Ideally, it should capture the full area and as large as possible a solid angle of the light emitted from the IR source. All the collected light should be focused on the sample and subsequently onto the detector. In our case, the commercial spectrometer sets the boundary conditions for the source (area and light collection), while the existing UHV setup determines the desired size of the illuminated surface area, which should be the adsorbate-covered sample area (3.5 mm molecular beam spot diameter[27]). The layout of the UHV system also imposes constraints on the mirror arrangement. Étendue was then used to estimate the position and shape of optical components to optimize the illumination on the sample and find a good detector position.

The focal length of the illumination and collection mirror must be short enough to achieve the large solid angle required for a small, illuminated spot. These considerations provide the basis for the final optimization performed by standard numerical ray tracing, which resulted in a short focal length of the illumination mirror of only a few centimeters, leading to a mirror arrangement with two mirrors in the UHV chamber. Placing mirrors outside of the UHV chamber would require a much larger focal length for the illumination mirror, accompanied by either unpractically large diameters of the mirrors and windows for the IR light or a loss of throughput by limiting the solid angle. The short focal length of the sample-facing mirrors is one of the key features characterizing our design; other systems usually use longer focal lengths.[20,40–44] In addition, low-loss light transfer from the spectrometer is achieved using elliptical mirrors to overcome the distance of almost one meter between the spectrometer and the sample. Furthermore, the ray tracing optimization took manufacturing tolerances into account to avoid a design highly sensitive to small deviations from the nominal geometry. These optimization elements are essential in guaranteeing a high-performance IRAS system.

A schematic diagram of the IRAS system is shown in FIG. 3. The illumination path features three mirrors directing the IR light from the spectrometer exit to the sample focus. After the almost parallel beam exits the spectrometer (Bruker VERTEX 80v) through a wedged $BaF_2$ window (3.5 mm thick), the light is focused into the polarizer focus by an off-axis parabolic mirror ('spectrometer-link mirror' in FIG. 3). A rotatable holographic wire-grid polarizer is placed after the focus point. The input mirror, an off-axis elliptical mirror, transfers the light through the shaping focus, a BaF$_2$ window (3 mm thick), and two angle-selection plates to the illumination mirror. This off-axis elliptical mirror features a high numerical aperture, illuminating the sample with a cone angle of 39° due to its relatively short focal length. As a result, the illumination mirror creates a small, near-circular illumination spot when the IR beam is directed at the sample at normal incidence.

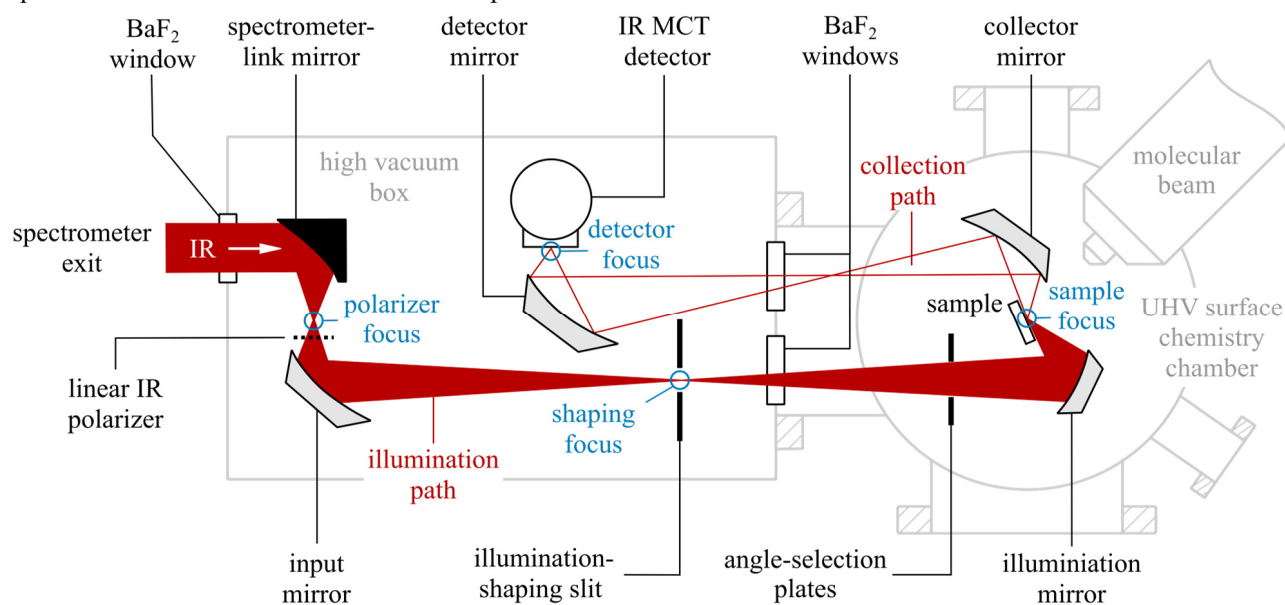

FIG. 3. Schematic top view of the IRAS system optics. The main optical components and focus points along the illumination (full red) and collection beam paths (red outline) are shown. Elliptical mirrors are filled in grey, and parabolic mirrors are black. BaF$_2$ windows separate chambers with different pressure levels. The components are not shown to scale and are simplified to ensure clarity.

As mentioned above, a small illumination spot on the sample is essential to reflect all light from the sample surface and keep most of the intensity focused on the molecular beam spot. The presence of a focus (shaping focus) between the input mirror and the illumination mirror facilitates low-loss light transfer over a distance of almost one meter. An aperture called the illumination-shaping slit is placed directly in the shaping focus; its task is to limit the illuminated area on the surface, especially at rather grazing incidence angles (see section III.B.2). In addition, the two angle-selection plates located between the shaping focus and the illumination mirror control the illumination angle range onto the sample. The specific function of the illumination-shaping slit and the angle-selection plates will be explained in section III.B.

The ideal sample illumination spot size is primarily defined by the adsorbate-covered area on the sample. Independent of the single crystal's surface area, the infrared beam should be restricted to the adsorbate-covered area created by the molecular beam, i.e., a circular region with a diameter of 3.5 mm in the center of the sample surface. Consequently, the measurement should include only radiation reflected from this region. IR radiation reflected from the uncovered sample regions contributes only to the noise, not to the signal. This consideration is also valid for a grazing incidence of radiation. Since the principal ray of the IR beam is not perpendicular to the sample surface, the roughly circular focus gets distorted into an elliptical shape. The major axis of the ellipse is oriented along the incidence plane, and the minor axis remains unchanged and equals the diameter of the focal spot. Hence, for very grazing incidence, the spot size is larger than the sample and adsorbate area, resulting in decreased sensitivity.

The IR light reflected from the sample is collected in the beam path drawn with a red outline in FIG. 3. This path features two elliptical mirrors, one wedged BaF$_2$ window in between, and the IR detector. The defining aspects of this part of the IRAS system are the complete collection of specularly reflected light and subsequent transfer to the detector with minimal losses. To collect all the light reflected from the sample, the collector mirror features a slightly shorter front focal length than the back focal length of the illumination mirror. This makes the setup insensitive to angular errors originating from the sample tilt. Since the sensitive detector area (1×1 mm$^2$) is smaller than the illuminated area on the sample, conservation of étendue requires a very large solid angle of the light entering the detector. The focal lengths of the collector mirror and the detector mirror are balanced to fit the detector light acceptance angle and the element size of the detector within the spatial restrictions of the measurement chamber to ensure maximum throughput. Additionally, the intermediate focal point between the collector mirror and detector mirror minimizes losses compared to using a parallel beam, overcoming a distance of approximately 0.6 m between the sample and detector.

Besides the conservation of étendue, the optical throughput is also defined by the transmissive and reflective losses of the optical components. Therefore, the correct selection of materials for optical components for the targeted IR wavenumber range and reducing the optical surfaces in the IRAS system is vital. To minimize reflective losses, all four off-axis elliptical mirrors (grey in FIG. 3) are manufactured from EN AW-6061-T6 aluminum, which has a reflectivity close to unity in the mid-IR region. These mirrors were fabricated according to custom specifications (for the focal lengths, see TAB. SI in the Supplemental Information), and the reflective surface is only covered by the native oxide layer of aluminum. The spectrometer-link mirror (black in FIG. 3) is a commercial gold-plated, off-axis parabolic mirror. Three BaF$_2$ windows are used to separate the vacuum chambers (grey) with different pressure regimes: FTIR spectrometer (~1 mbar), high-vacuum box (~10$^{-3}$ mbar), and UHV surface chemistry chamber (~5×10$^{-11}$ mbar), where the sample is located. The BaF$_2$ window thickness has to be chosen as a compromise between mechanical stability and optical losses near the distinct cutoff in the transmission of BaF$_2$ at approximately 1000 cm$^{-1}$. All windows feature a half-degree wedge angle to avoid interference peaks. Also, the linear holographic polarizer uses BaF$_2$ as substrate material. Our system utilizes a highly sensitive liquid-nitrogen-cooled mercury cadmium telluride (MCT) detector from InfraRed Associates with a 1×1 mm$^2$ detector element and a field view of 60°. This detector features a BaF$_2$ window, a spectral range of 850 cm$^{-1}$ to 12000 cm$^{-1}$, a specific detectivity bigger than 4×10$^{10}$ cm Hz$^{1/2}$ W$^{-1}$, and a liquid nitrogen dewar lasting 12 h.

### B. Peak Height Optimization

In addition to maximizing throughput, optimizing the peak heights was a central goal of our design strategy. Careful balancing of these measures is crucial, as optimizing peak heights at the cost of decreased intensity can improve the SNR. Maximizing peak heights requires restricting illumination to the adsorbate-covered area, clean polarization of the IR beam, and a precise adjustment of the incidence angle range by the angle-selection plates (see FIG. 4). The importance of selecting the incidence angles as laid out in section II, and is an innovation in the surface science community.

#### 1. Effect of the Incidence Angle θ

The choice of the incidence angle range depends on the polarization, the orientation of the dipole moment of the adsorbate, and the Brewster angle of the investigated dielectric.

The IRAS system described here offers a wide range of incidence angles from 48° to 87°. We have implemented angle-selection plates that can limit the range of incidence angles at either side (see FIG. 3 and FIG. 4). The main application of the angle-selection plates is for p-polarized light, where the opposite sign of the peaks above and below the Brewster angle $\theta_B$ would lead to a cancellation of the total signal (see FIG. 2); this can be avoided by selecting only angles at one side of $\theta_B$. Since $\theta_B$ depends on the refractive index of the substrate in use, our setup allows variation of the cutoff angle $\theta_L$ to optimize the IR measurement for different samples. Depending on the measurement system, the variation of $\theta_{min}$ and $\theta_{max}$ additionally to $\theta_L$ can lead to SNR improvements. For s-polarization, the complete throughput of the system can be utilized without restricting the incidence angle range.

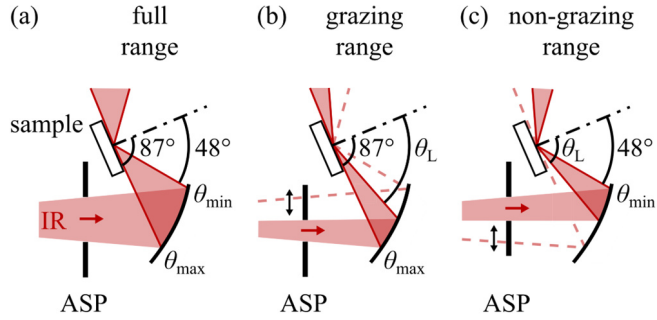

FIG. 4. Schematic function of the angle-selection plates (ASP) in three main measurement configurations. The plates are located before the illumination mirror, reducing the incidence angle range. Without using the ASP, the incidence angles are between $\theta_{min}$ = 48° and $\theta_{max}$ = 87°. (a) shows the setting for the full angle range, (b) selection of grazing incidence, and (c) selection of non-grazing incidence angles.

In our IRAS setup, the angle-selection plates are placed in the UHV chamber before the illumination mirror (see FIG. 3). The two plates can be moved independently, which makes it possible to select both the minimum and the maximum incidence angle. In most cases, only one of the plates will be used at a time. Section V.B provides measured data for the non-grazing, grazing, and full angle range.

### 2. *Effects on Peak Height in the Real System*

A realistic view of an IRAS system must take into account that, e.g., imperfect optical components or a finite source will alter the IR beam characteristics compared to the ideal case. Effects like the degree of light polarization,[45] the incidence angle spread,[24,36] or the light reflected from the adsorbate-free area affect the maximum achievable peak height.

The degree of light polarization is limited by the polarizer itself and polarization aberrations caused by rays that are not in the plane of incidence of the principal ray and reflected from the mirror surfaces (skew aberration).[45–47] The depolarization effect on the peak heights can be estimated from a weighted sum of the reflectivities of the samples[45] for the two polarization directions, p- and s-polarization. Depending on the degree of depolarization, signals from p-polarization may leak into the s-polarization when the incidence angle range is sufficiently far from the Brewster angle. Usually, the opposite effect, leakage of s-polarized light into measurements with p-polarization, is more relevant: The zero of reflectivity at the Brewster angle will disappear when a small s-component is present, and thus, the singularity in FIG. 2 will disappear. A distribution of incidence angles[24,36] will also smear the singularity.

Because our setup features a small adsorbate-covered area, the magnitude of the $\Delta R/R_0$ ratio will also be reduced by rays reflected from uncovered areas. We deposit molecules with a molecular beam,[27] creating a 3.5 mm circular adsorbate-covered region in the sample center, the molecular beam spot. FIG. 5 (a) shows a photograph of the sample holder, including a mounted sample. The dark circular area is the molecular beam spot. At grazing incidence, the illumination spot (and the collected light at the detector) includes regions outside the molecular beam spot (vicinity), as shown by the ray tracing simulation in FIG. 5 (b). For a given angle of incidence, the reflected light from the vicinity reduces the peak height. To reduce the intensity of the vicinity illumination, we implemented the illumination-shaping slit in our system (see FIG. 3). It consists of two fixed aperture blades that modify the illumination on the sample to almost rectangular (see FIG. 9). For incidence angles below 65°, almost all light at the detector comes from within the measurement area is concentrated at the MB spot. At more grazing angles (>65°), the vicinity is illuminated; up to ≈48% of the detected radiation stems from the adsorbate-free region. In case all of the sample area is covered by the adsorbate (background dosing), a large illuminated area poses no limitation, and one may also omit the illumination-shaping slit shown in FIG. *3*.

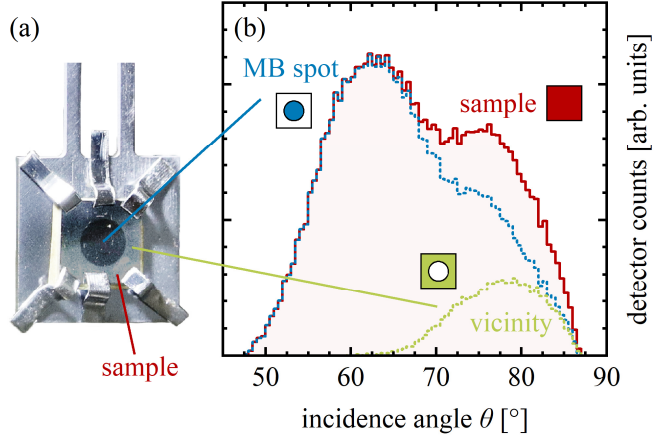

FIG. 5. Molecular beam (MB) on the sample. (a) shows a photograph of the sample holder, including the sample (6×6 mm$^2$ surface area). (b) Simulated intensity at the detector resulting from reflection from the different areas of the sample surface as a function of the incidence angle on the sample surface. The red curve represents the total intensity on the detector reflected from the sample. The blue dotted curve shows the intensity reflected inside the molecular beam spot. The green dotted curve represents the intensity reflected from the vicinity (region not covered by adsorbates) around the molecular beam spot.

FIG. 6. displays the calculated p-polarized $\Delta R/R_0$ for TiO$_2$ at $\tilde{\nu}$ = 2178 cm$^{-1}$ as a function of the incidence angle angle $\theta$ on the sample, including the effects of incidence angle spread, incomplete polarization, and a partly adsorbate-free sample. The calculation was performed for a model adsorbate with a dipole moment perpendicular to the surface, and the collection efficiency was taken into account. Curve (4) includes a ±2° angular spread in the angle of incidence, curve (3) shows the effect of the finite efficiency of the polarizer (99.5%), and curve (2) demonstrates the effect of the light that gets reflected from the adsorbate-free area of the sample in our system. The effects of the angular spread perpendicular to the angle of incidence and depolarization by the mirrors (skew aberration) are not included in this figure. The strongest effect near the Brewster angle comes from the 0.5% leakage of s-polarized light, whereas the deterioration of the $\Delta R/R_0$ ratio at grazing incidence is dominated by light reflected outside the adsorbate-covered area. This is expected since it becomes increasingly difficult to focus all light into the adsorbate-covered area at very grazing angles. The same considerations are also valid for p-polarization and in-plane dipole moments (FIG. 5b). For IRAS measurements with s-polarization, leakage of p-polarized light is irrelevant in the region around the Brewster angle. At grazing angles, the relative peak height will be reduced by reflection from outside the adsorbate-covered area for p and s-polarization (see FIG. 5b).

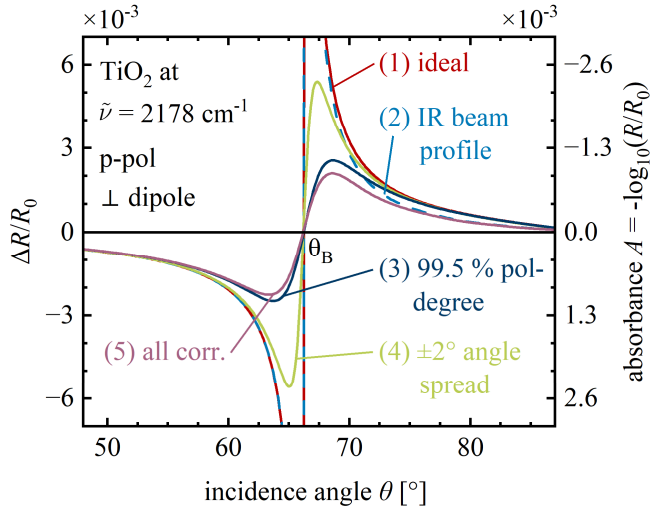

FIG. 6. Different effects influencing the calculated normalized reflectivity difference $\Delta R/R_0$ for a model adsorbate with vertical dipole moment on TiO$_2$. The calculation was performed for p-polarization. Curve (1) shows the uncorrected case, and curves (2)–(4) depict the separate effects of (2) illumination of sample areas not covered by the adsorbate, (3) a polarization degree of 99.5%, and (4) a ±2° spread of the incidence angle. Curve (5) shows a calculation including all effects (2)–(4).

Due to the higher reflectivity of the sample for s-polarized light, compared to p-polarized light, especially around the Brewster angle (see FIG. 1), it is primarily p-polarized spectra that suffer from the depolarization effect. The corrected curves for p-polarization and s-polarization with parallel dipole moment orientations can be seen in FIG. S1 in the Supplementary Information.

However, depolarization and IR beam divergence significantly decrease the normalized reflectivity difference around $\theta_B$, resulting in lower peak heights in the final spectrum. As exemplified here for $TiO_2$ in FIG. 1, the vanishing reflectivity at the Brewster angle pronounces the described divergence effect and depolarization in materials with an almost zero extinction coefficient $k$.

## IV. MECHANICAL DESIGN - REALIZATION

In addition to optimization of the optical design, it is also essential to achieve mechanical stability of the setup. This reduces the sensitivity to environmental vibrations (e.g., vacuum pumps) and is also required to maintain the alignment of the optics and ensure precise and reproducible sample positioning. Factors such as temperature stability of the laboratory, sample vibrations, or tilt errors[48] of the sample mounted on a cryostat influence the spectral stability, i.e., equal baselines for consecutive measurements. The setup is designed to fit an already existing UHV measurement chamber,[49] which takes these considerations into account. It is characterized by precise positioning of the optical components, minimization of adjustable components (kinematic mounts), and a highly stable optical platform. All mirror holders feature precise pinned fittings and spring-loaded screws to keep them securely in place.

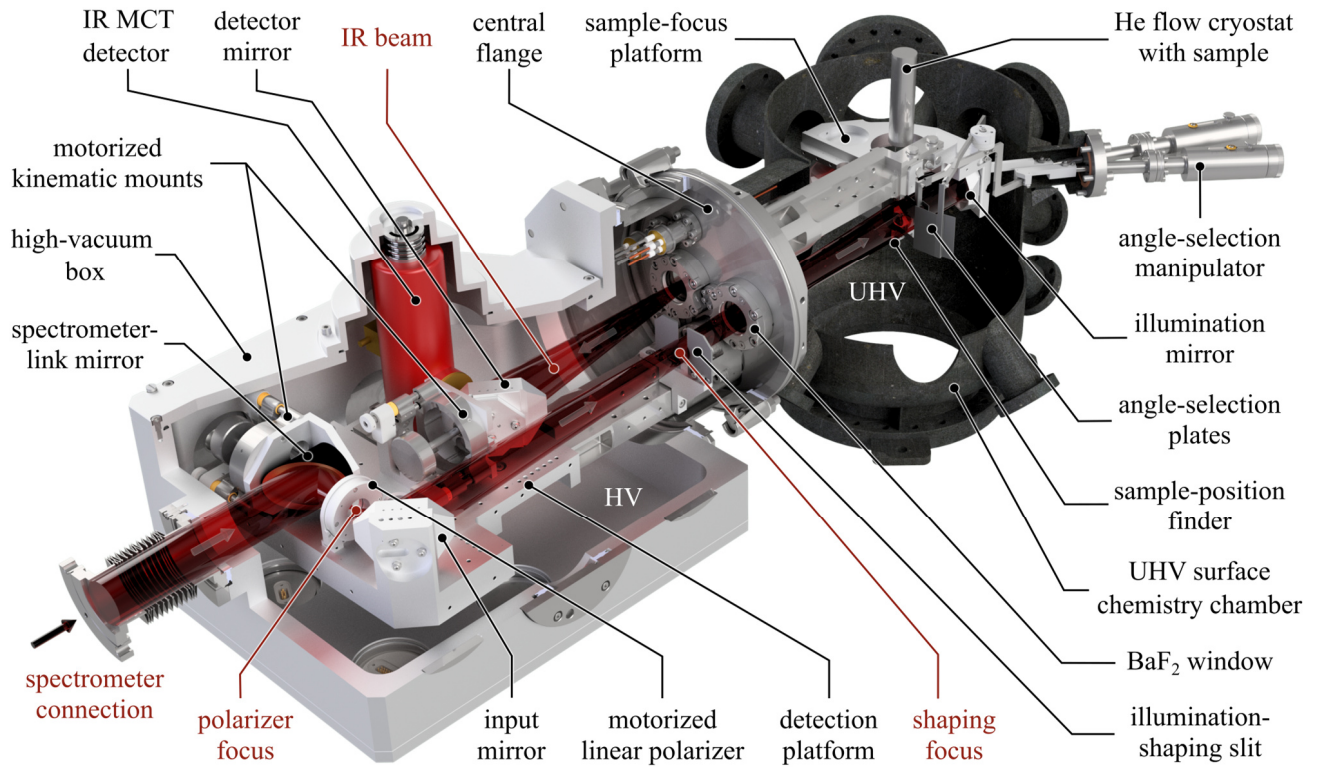

FIG. 7. Partial section view of the key components of the IRAS setup. The IR beam path is visualized in red. The arrows in the beam indicate the direction of the IR light. The UHV surface chemistry chamber is shown in black. The detection platform is in the high vacuum box. The sample-focus platform and the angle-selection plates are inside the UHV chamber. The sample is mounted on the bottom of a helium flow cryostat and located in the IRAS measurement position, but it is hidden behind the sample-focus platform.

FIG. 7 shows a partial section view of the IRAS setup with its primary components. The IRAS setup connects a commercial FTIR spectrometer (Bruker VERTEX 80v; exit window on the left side of FIG. 7) and a UHV surface chemistry chamber[49] (black chamber on the right side). The detection platform with the transfer optics, polarizer, and detector are inside the high-vacuum box, which is pumped to a pressure of $1.5 \times 10^{-3}$ mbar. The high-vacuum box is machined from EN AW-6061-T6 aluminum; two covers and Viton gaskets seal its top and bottom. The top cover includes a liquid nitrogen filling port for the IR MCT detector, and the bottom cover feedthroughs for electrical connections. The right side of the HV box is connected with a DN 200 ISO-K tube to the central flange of the IRAS system.

All optical components in the high-vacuum box and the UHV chamber are supported by this central flange, which provides precise positioning. The stability of the system depends sensitively on this link. Milled corner joints are used to ensure accurate and reproducible connections between the central flange of the IR setup and rigid customized stainless steel supports that hold the detection platform in the high-vacuum box and the sample-focus platform on the UHV side. The central flange includes two wedged $BaF_2$ windows for entry and exit of the IR beam into the UHV chamber. These are each fixed on a custom CF-40 flange to facilitate exchange if necessary. Thermocouple feedthroughs are mounted on the flange for temperature monitoring during bakeout.

The optical components outside the FTIR spectrometer are located on the two platforms. The detection platform on the high-vacuum side has two 2-axis motorized kinematic mounts (New Focus 8822-AC-UHV) for IR beam alignment (with the spectrometer-link mirror) on the illumination and detector side (with the detector mirror). The input mirror is attached rigidly to the detection platform. The polarizer is mounted on a motorized rotation stage (SR-5714, SmarAct GmbH). With a precision of 0.1° or better, the rotation angle uncertainty of the stage does not contribute to the depolarization of the beam.

The MCT detector, the red cylinder in FIG. 7, is located on a homemade mount and is manually adjustable in the x, y, and z directions. It allows for slight tilting adjustments around all axes. After the initial alignment, the position of the detector is determined by a milled corner joint, ensuring reproducible remounting. The detector electronics supplied by Bruker was removed from the bottom of the detector and placed outside the high-vacuum box, with the wiring (coaxial cable) passing via a current feedthrough to the detector.

The detection platform itself is manufactured from aluminum EN AW-6061-T6 and mounted to the stainless steel support using a milled corner joint and a spring-loaded screw connection.

The two sample-side mirrors in UHV are rigidly mounted with pins and spring-loaded screws on the sample-focus platform. A hole in the center of the sample-focus platform allows the sample and the cryostat to pass to the measurement position. It provides room for adjustments in x-, y-, and z-directions and rotation around the z-axis through a sample manipulator. The hole diameter (60 mm) is small enough to avoid the deposition of sputter debris on the mirrors when the sample is in the sputtering position (200 mm above the IRAS and molecular beam position). Cu shields (not shown in FIG. 7) protect the mirrors and IR windows from material deposited with the electron-beam evaporators below the IRAS components in the UHV system.

The angle-selection plates are located in UHV on the side of the illumination beam. The rotation mechanism of each angle-selection plate is operated by a linear motion feedthrough (angle-selection manipulators in FIG. 7). The position of each plate can be read from a scale engraved in the rotation mechanism; these values directly translate into the cutoff angle for the incident beam.

A sample-position finder (see FIG. 8) mounted on the lower part of the sample-focus platform (visible in FIG. 7 behind the IR beam) enables the user to see the sample from the bottom, overlaid with a scale grid. This facilitates quick and reproducible sample positioning.

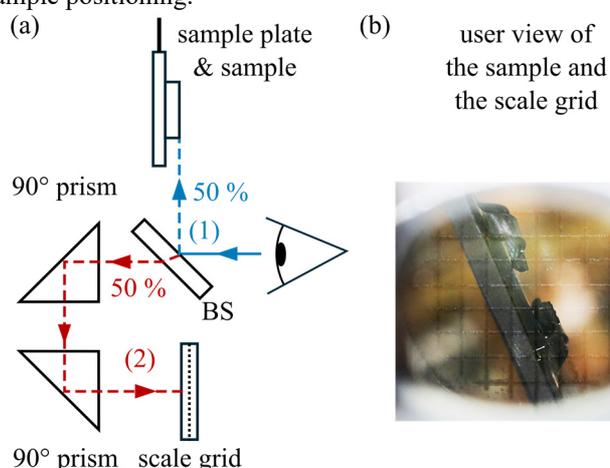

FIG. 8. Function of the sample-position finder. (a) shows two schematic optical paths, (1) and (2), separated by the 50:50 beam splitter (BS) from Thorlabs (BSW04). Both paths have the same optical path length. Path (1) reflects on the BS to the sample. Path (2) goes through the BS, reflected by two 90° fused silica prisms and guided to the scale grid. (b) shows the resulting image that is visible to the user. The sample bottom and the scale grid with a 1 mm grid spacing can be seen simultaneously.

During the bakeout of the UHV chamber, the IRAS system remains connected to the chamber. To avoid overheating of the components, the stainless steel beam holding the detection platform is connected via copper braids to an aluminum flange (on the bottom cover), where a water-cooling unit is attached. Therefore, no realignment is necessary after the bakeout. If optical realignment is required, the motorized kinematic mounts facilitate alignment without venting the high-vacuum box. These measures result in a mechanically highly stable system that achieves sensitive measurements with a flat baseline over long periods and benefit user-friendliness.

## V. PERFORMANCE ASSESSMENT

An initial performance characterization of our IRAS system was first carried out outside our UHV surface analysis chamber. The intensity distribution was measured at critical focal points of the system and compared to simulation data. Next, we tested the performance of the setup in UHV in comparison to IRAS results published in the literature.[12,14,20,22,50,51] The FTIR spectrometer Bruker VERTEX 80v with the standard mid-IR source (glowbar) was used for all the characterizations.

### A. Intensity Distributions Along the IR Beam Path

To measure the intensity distribution in the critical focal points, the spectrometer was connected to a home-built optical table, with the IRAS setup mounted and exposed to atmospheric conditions. The beam-defining aperture (J-stop) inside the spectrometer was 6 mm during these measurements. The moving interferometer mirror of the FTIR spectrometer was stopped (usually, it is in motion during IR measurements), and the interferometer laser was switched off to prevent saturating the camera sensor by the laser beam.

The angle-selection plates were completely open, not obstructing the IR beam. The intensity characterization was performed without the $BaF_2$ windows mounted on the central flange and without the polarizer. Custom-made holders positioned the sensor in the sample and detector focal points, with the position according to the simulation model. A flat aluminum mirror (Ø 12.6 mm) replaced the sample for the measurements in the detector position.

To measure the intensity maps, we utilized the CMOS camera DMM 37UX178-ML from The Imaging Source. The sensitivity of its sensor chip (Sony IMX 178, back-illuminated) extends into the near-infrared region. A thin, framed glass plate covers the sensor, whose pixels exhibit a limited light acceptance angle. Therefore, the intensity maps do not include light hitting the sensor at grazing angles.

FIG. 9 shows the measured and simulated intensity distribution at the sample and detector focal points. The two intensity maps in the sample focus (FIG. 9 (a) and (c)) show good agreement in the central region (red and yellow). The experimental image indicates slightly better focusing than in the simulation (smaller FWHM values). We attribute this to the limited acceptance angle of the camera, which suppresses light at grazing incidence, where the focus area is larger.
In the detector position (FIG. 9 (b) and (d)), we find excellent agreement between the measured and simulated intensity distributions in the high-intensity region; also, the FWHM values agree very well. The low-intensity tails of the distribution are more pronounced in the experiment than in the simulation. In summary, the intensity maps verify the correct simulations and design of the spectrometer setup.

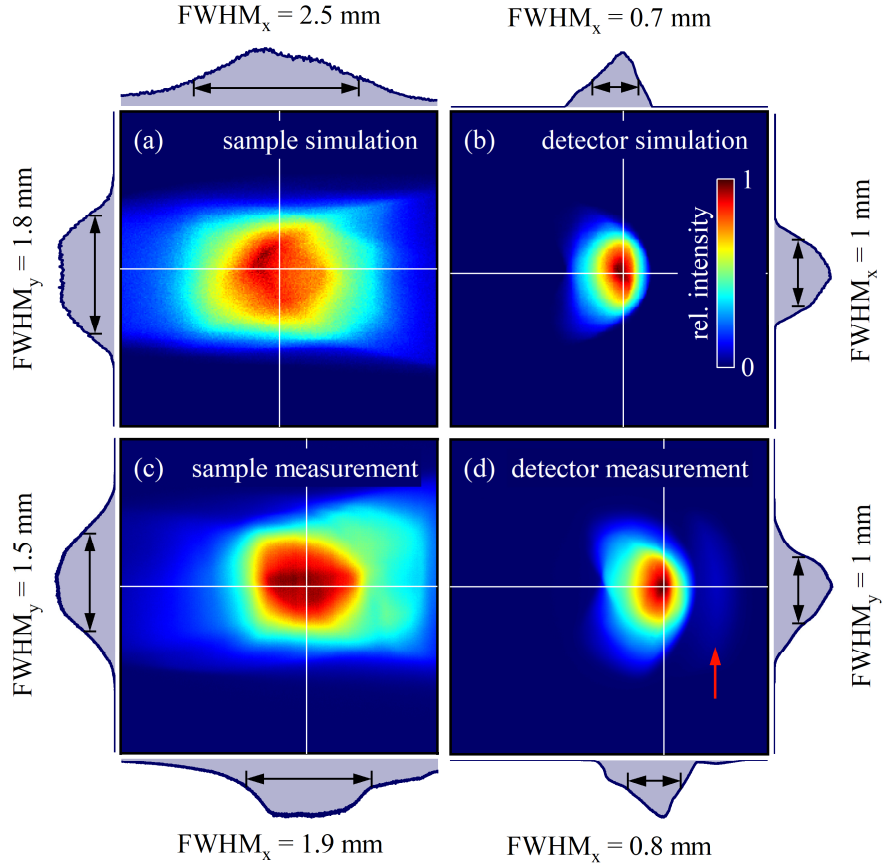

FIG. 9. Simulated (a, b) and measured (c, d) intensity distributions in two positions of the IRAS setup: The sample position (left) and the position of the detector (right). The white lines in the maps cross the intensity-weighted centroid ("center of mass") and indicate the position of the line profiles displayed at the periphery of the figure. The black double arrows in the line profile plots indicate the full width at half maximum (FWHM). The red arrow marks an additional feature originating from the mirror mount.

### B. Spectroscopic Performance for CO and $D_2O$ on $TiO_2(110)$

The IRAS setup was tested using a synthetic rutile $TiO_2(110)$ single crystal (CrysTec GmbH, miscut < 0.05°) mounted on a Ta plate (with Ta clips) and cooled with a helium flow cryostat (base temperature of 37 K) of the UHV setup.[49] The UHV chamber features a base pressure of $5\times10^{-11}$ mbar. For good thermal contact, a gold foil (0.025 mm thickness) was placed between the $TiO_2$ crystal and the Ta plate. The [001] direction of the $TiO_2(110)$ sample was in the incidence plane of the IR beam.

The sample was first annealed to 950 K. After that, the sample was prepared by cycles of sputtering (1 keV $Ne^+$, 15 min, at 300 K) and UHV annealing to 900 K for 20 min. From time to time, the sample was reoxidized by annealing at 900 K in $5\times10^{-7}$ mbar $O_2$ and subsequent UHV annealing to 900 K for 10 min.[41,52,53] The reduction state of the sample is not necessarily the same in all measurements presented here; it could have changed because of the repeated preparation steps during the lengthy test measurements.

IR spectra were recorded with a mirror speed of 19 mm/s (laser interferometer frequency 60 kHz), a zero filling factor of one, and a Happ-Genzl apodization. For p-polarized spectra, the spectrometer was set to average 1000 scans with a resolution of 4 $cm^{-1}$ and a J-stop of 6 mm unless mentioned otherwise. Thus, an IRAS measurement took about five minutes in total for the reference and sample spectrum. S-polarized measurements required a smaller J-stop setting of 3 mm to avoid saturation of the MCT detector. Again, a resolution of 4 $cm^{-1}$ was used; averaging was done over 4000 scans (≈ 20 minutes).

Usually, IRAS spectra require the recording of two spectra – the reference spectrum ($R_0$) and the sample spectrum ($R$) – which are then used to calculate the normalized reflectivity difference according to equation (1). Here, the IR spectra were measured using two different approaches.

The first approach is to measure the clean sample for a reference spectrum. Then, the sample is moved to the molecular beam position for dosing. After moving the sample back to the IR position, the measurement from the adsorbate-covered surface is used as a sample spectrum. This method will be referred to as the adsorption spectrum and describes the standard procedure for IRAS measurements.

For a smooth workflow, the reference spectrum of the CO measurements was sometimes recorded after measuring the adsorbate-covered surface. Then, the adsorbate was desorbed at 350 K, and the reference spectrum (after the base temperature of 43.5 K was reached) was acquired. Spectra acquired in this manner will be referred to as desorption spectra in the following. This procedure has the advantage that no sample movement is necessary between the acquisition of the sample and reference spectra, which results in a better match between the sample positions for these two spectra.

The SNR of the normalized reflectivity difference spectra was evaluated by determining the peak heights and the root mean square (RMS) noise of the spectra in the range from 1900 cm$^{-1}$ to 2100 cm$^{-1}$ using the OPUS software[54] and a parabolic fit.

### 1. *CO on Rutile TiO$_2$(110)*

Low coverages of CO adsorbed on rutile TiO$_2$(110) provide a good model system for checking the functionality of our IRAS setup for vibrational modes perpendicular to the surface, because previous studies[14,20,22,50,51] provided comprehensive IR measurements for this system.

#### a. *Angle Range Optimization*

To optimize our IRAS measurements, we need to know how the SNR changes with different incidence angle ranges. Adjusting the angle-selection plates (see FIG. 3 and FIG. 4) allows us to change the incidence angles in the non-grazing and grazing range. For the measurements presented here, the errors of the incidence angle are estimated to be within ±5° due to limited precision in sample positioning, beam divergence, and uncertainties in the calibration of the angle scales of the plates. All data in this section were obtained with a coverage of 1 monolayer (ML) CO, which corresponds to a gas dose of 1.36 L (Langmuir; 1 L = 10$^{-6}$ Torr s) at a temperature of 43.5 K. Since the angle-selection plates were moved in each step, each measurement also includes a reference spectrum for this range of incidence angles (taken after desorption).

Comparing the non-grazing and grazing sides of the Brewster angle, we find the optimum SNR at the non-grazing side, which can be seen by directly comparing the spectra in FIG. *10* (b) and (c). Finding the optimum SNR at the non-grazing side is expected. The width of the incidence angle ranges between the limits and the Brewster angle for the two cases are comparable (18 and 21°). However, the non-grazing side provides two advantages: (i) a higher average signal intensity and (ii) in this range of incidence angles, all the beam gets reflected from the adsorbate-covered area of the sample in our setup (FIG. 5). The situation would be different for a material with a substantially lower index of refraction *n*, where the Brewster angle is lower (e.g., $\theta_B = 56°$ for $n = 1.5$). In such a case, the angle range at the non-grazing side will be very small, and the best SNR will be obtained at the grazing side of $\theta_B$.

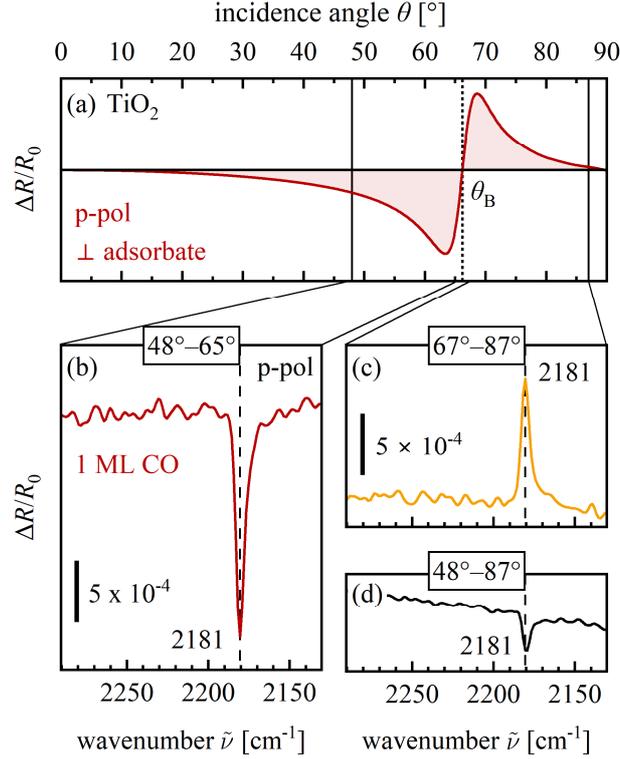

FIG. 10. Experimental spectra for 1 ML CO/TiO$_2$(110) obtained for different incidence angle ranges. Frame (a) shows the calculated angle-dependence of $\Delta R/R_0$ for comparison. Note the low peak height of spectrum (d), which results from partial cancellation of the positive and negative peak heights on the right and left side of the Brewster angle. Thus, in spite of the high intensity leading to low noise, the SNR of this measurement is only 24.4, much less than in (b) or (c) (SNR = 69.4 and 51.3, respectively) where only peaks with the same orientation contribute to the signal. The spectra were measured with a resolution of 4 cm$^{-1}$ and averaged over 1000 scans. No baseline correction was applied, and all spectra are plotted with the same scale.

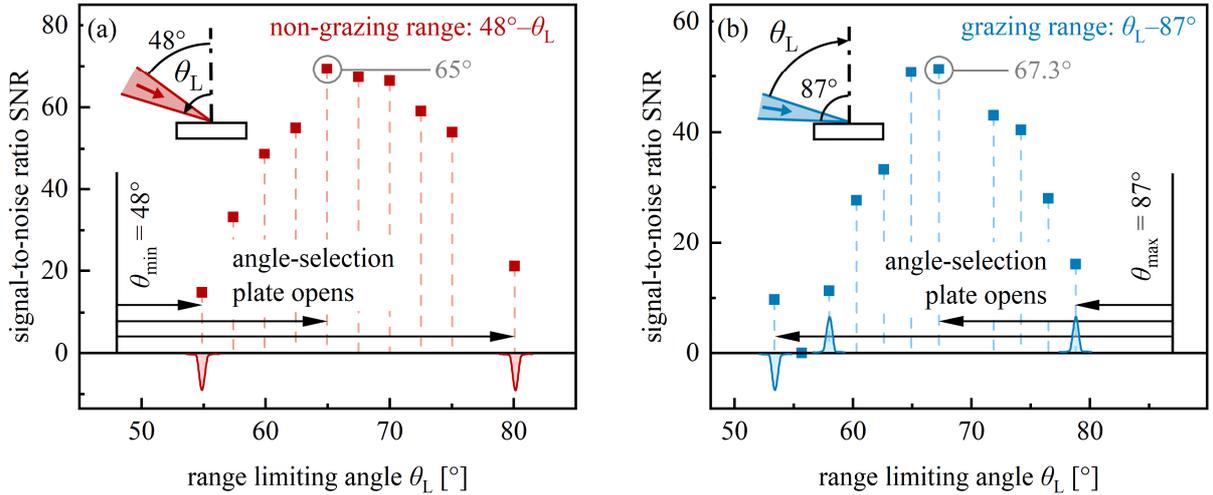

FIG. 11. SNR for CO/TiO$_2$(110) (vertical dipole moment) and different non-grazing and grazing incidence angle ranges. A grey circle marks the maximum SNR. For each plot, one angle-selection plate is fixed (indicated by the vertical black line) not to obstruct the IR beam, and the second plate is moved to limit the incidence angle by $\theta_L$ (dashed lines). See FIG. 4 for the measurement configuration. The schematic peaks on the zero line indicate the orientation of the peaks in the $\Delta R/R_0$ spectrum at the given angle range.

FIG. *11* demonstrates the impact of the incidence angle range on the SNR. Plot (a) depicts the SNR measurements with non-grazing angles between 48° and $\theta_L$ for different $\theta_L$ values. With increasing $\theta_L$, i.e., increasing slit size (indicated by the black arrows), more light passes through, and the SNR increases. The maximum SNR of 69.4 is reached at $\theta_L = 65°$ (see FIG. 10 (b)). This is close to the Brewster angle of 66.2°. The SNR decreases when the plate is opened more towards the grazing side. This can be directly related to the band inversion above the Brewster angle: When including angles at both sides of $\theta_B$, positive and negative peaks superimpose and partly cancel out, leading to a decrease in the SNR although the intensity increases (see also FIG. 2).

In FIG. 11 (b), the SNR for the grazing range between $\theta_L$ and $\theta_{max} = 87°$ is plotted. Again, the SNR increases with increasing intensity as the plate at the non-grazing side opens. The maximum SNR is reached at $\theta_L = 67.3°$ (see FIG. 10 (c)), again close to the Brewster angle. In the spectra recorded, the peak orientation changes at a cutoff angle of $\theta_L = 56°$. Here, the positive and negative contributions cancel out; thus, the SNR is zero.

Finally, we should mention that the best SNR is obtained for incidence angle ranges not reaching to the Brewster angle $\theta_B$. Although the difference between the optimal cutoff $\theta_L$ and the Brewster angle is within the error bars, we consider this a valid result. In a real system, the $\Delta R/R_0$ ratio close to the Brewster angle is low (see FIG. 6). The reflected light close to the Brewster angle will be dominated by s-polarization (leaking through the polarizer or caused by polarization aberration); this light carries no signal, but it can contribute to the noise and decrease the peak height in the $\Delta R/R_0$ spectrum.

b. *Spectra at High Resolution and Low Coverage*

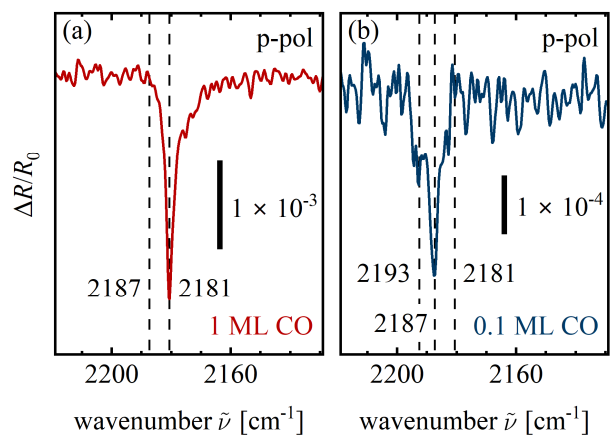

FIG 12. Raw spectra for (a) 1 ML and (b) 0.1 ML CO/TiO$_2$(110) acquired with 1 cm$^{-1}$ resolution and an incidence angle range of 48° to 67°. Averaging was done over (a) 1000 and (b) 4000 repetitions.

FIG 12 shows two spectra acquired with a high resolution of 1 cm$^{-1}$ and the optimized incidence angle range (48°–67°). Plot (a) shows 1 ML CO adsorbed on the surface. The CO stretch again appears at 2181 cm$^{-1}$; the SNR obtained with 1000 repetitions is 43. FIG 12 (b) shows the spectrum obtained for 0.1 ML CO acquired with 4000 scans. The main peak appears at 2187 cm$^{-1}$ and has a SNR of 11.6. A side peak is visible at 2193 cm$^{-1}$; but no peak is seen at the position of the 1 ML peak (2181 cm$^{-1}$). The height of the main peak for 0.1 ML CO is seven times smaller than for 1 ML CO.

The peak positions and shapes generally agree with the work presented previously.[14,20] The small differences in the peak position (2181 cm$^{-1}$ in our work vs. 2178 cm$^{-1}$ in the literature for 1 ML CO) may be due to a different reduction state of the TiO$_2$(110) sample. Detailed parameters of the shown spectra can be seen in Table below.

### 2. 1 ML D$_2$O on Rutile TiO$_2$(110)

The performance in s-polarized IRAS measurements was assessed by adsorbing 1 ML (1.15 L) D$_2$O on TiO$_2$(110) and utilizing the full incidence angle range. The vibrational modes in this system are mainly oriented parallel to the surface. This system has also been well studied[12,22] and provides an excellent foundation for benchmarking our system.

The sample was prepared using the standard procedure described in section V.B. After flashing it to 950 K, it was cooled to 45.9 K, and a reference spectrum was acquired. After hydroxylation (through reaction with the oxygen vacancies) at 315.4 K with 1.5 L of D$_2$O, the reduced TiO$_2$(110) sample was further treated by adsorbing 1 ML D$_2$O at 186.4 K. Following this, a sample spectrum was recorded at 45.9 K with a resolution of 4 cm$^{-1}$ and 4000 repetitions (FIG *13*). It shows two peaks. The first is located at 2602 cm$^{-1}$, and the second, broader feature with the maximum at 2328 cm$^{-1}$, has a SNR of 64.5. The spectrum shape generally agrees with the spectra reported in previous work[22] and was measured in approximately 20 minutes.

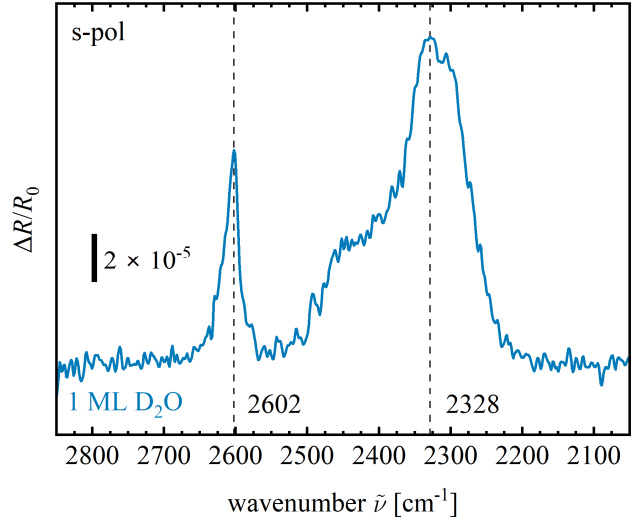

FIG 13. S-polarized spectra of 1 ML D$_2$O on the hydroxylated TiO$_2$(110) surface with the [001] direction in the incidence plane. The spectrum was acquired with 4 cm$^{-1}$ resolution and averaging was done over 4000 repetitions. A baseline correction was applied to the spectrum.

| resolution [cm$^{-1}$] | # scans | adsorbate | angle range | pol. | RMS noise | peak height | SNR |
|---|---|---|---|---|---|---|---|
| 4 | 1000 | 1 ML CO | 48°–87° | p | 1.23×10$^{-5}$ | 3.00×10$^{-4}$ | 24.4 |
| 4 | 1000 | 1 ML CO | 48°–65° | p | 2.63×10$^{-5}$ | 1.83×10$^{-3}$ | 69.4 |
| 4 | 1000 | 1 ML CO | 67°–87° | p | 2.02×10$^{-5}$ | 1.04×10$^{-3}$ | 51.3 |
| 1 | 4000 | 0.1 ML CO | 48°–65° | p | 3.01×10$^{-5}$ | 3.50×10$^{-4}$ | 11.6 |
| 1 | 1000 | 1 ML CO | 48°–65° | p | 5.74×10$^{-5}$ | 2.50×10$^{-3}$ | 43.6 |
| 4 | 4000 | 1 ML D$_2$O | 48°–87° | s | 2.17×10$^{-6}$ | 1.40×10$^{-4}$ | 64.5 |

Table I. Root-mean-square (RMS) noise values, peak heights (both given as a fraction of the intensity), and signal-to-noise ratios SNRs in the different spectra. All measurements were performed on TiO$_2$(110) and are shown above. The noise values were evaluated in the wavenumber range from 1900 cm$^{-1}$ to 2100 cm$^{-1}$ from the $\Delta R/R_0$ with the OPUS software[54] from Bruker and the use of a parabolic fit.

Table I summarizes the analysis results of the spectra shown above. The RMS noise, the peak height in the spectrum, and the SNR are shown for different coverages of CO and D$_2$O adsorbed on the reduced TiO$_2$(110). The best result for p-polarization was achieved in the angle range from 48°–65° for 1 ML CO. Measurements of D$_2$O with s-polarization show good SNR for the full incidence angle range from 48°–87°.

## VI. SUMMARY

This work introduces a novel IRAS setup tailored to study adsorbates on dielectrics such as metal oxide single-crystals under UHV conditions. The most important features for achieving a high signal-to-noise ratio in measurements of a small sample area are (i) high throughput based on a careful design of the optics and a large numerical aperture obtained by using mirrors with short focal lengths, (ii) the selection of the optimal incidence angle range to only acquire signals with the same orientation of the peak, resulting in maximized peak heights and (iii) the precise integration of the optical platforms and components in the IRAS system ensuring long-term stability of the optical alignment and flat baselines. By considering these features, we achieved a signal-to-noise ratio close to 70 at 4 cm$^{-1}$ resolution acquired in a time of 5 minutes for p polarized IRAS of one monolayer CO on TiO$_2$(110) and sub-monolayer sensitivity with minutes of measurement time.

## SUPPLEMENTARY MATERIAL

The supplementary material includes the formulas used to calculate the reflectivities of the clean and adsorbate-covered surfaces, as well as calculated curves for the s- and p-polarization of the incidence angle dependent $\Delta R/R_0$, taking into account non-ideal polarization, angle spread, and adsorbate illumination. Additionally, the focal lengths for the mirrors used in the IRAS setup are provided.

## ACKNOWLEDGEMENTS

This work was supported by the Austrian Science Fund (FWF) (Start-prize, project no. Y847-N20 and SFB TACO, project no. F81) and the European Research Council (ERC) (Consolidator Grant 'E-SAC', Grant Agreement No. 864628 and Advanced Research Grant 'WatFun', Grant Agreement No. 883395). The authors thank Rainer Gärtner and Herbert Schmidt for the excellent fabrication of the custom-designed components. M. E. acknowledges support by the MSCA action under the Horizon Europe Framework Program for action 101103731 — SCI-PHI.

## AUTHOR DECLACATION

*Conflict of Interest*

The authors have no conflicts to disclose.

*Author Contributions*

**David Rath**: Methodology (lead); Validation (lead); Formal analysis (lead); Investigation (lead); Data curation (lead); Writing – original draft (lead); Writing - Review & Editing (equal); Visualization (lead). **Vojtěch Mikerásek**: Methodology (supporting); Investigation (supporting); Visualization (supporting). **Chunlei Wang**: Investigation (supporting); Writing - Review & Editing (supporting). **Moritz Eder**: Investigation (supporting); Writing - Review & Editing (supporting). **Michael Schmid**: Conceptualization (equal); Methodology (equal); Validation (equal); Formal analysis (equal); Investigation (supporting); Writing - Review & Editing (equal); Supervision (equal). **Ulrike Diebold**: Conceptualization (equal); Methodology (supporting); Validation (supporting); Investigation (supporting); Resources (equal); Writing - Review & Editing (equal); Supervision (supporting); Project administration (supporting); Funding acquisition (equal). **Gareth S. Parkinson**: Conceptualization (lead); Methodology (supporting); Validation (supporting); Investigation (supporting); Resources (lead); Writing - Review & Editing (equal); Supervision (supporting); Project administration (lead); Funding acquisition (lead). **Jiří Pavelec**: Conceptualization (equal); Methodology (equal); Validation (equal); Investigation (equal); Data curation (supporting); Writing – original draft (equal); Writing - Review & Editing (equal); Supervision (lead); Project administration (equal).

## DATA AVAILABILITY

The data that support the findings of this study are available from the corresponding author upon reasonable request.

# Supporting Information: Infrared Reflection Absorption Spectroscopy Setup with Incidence Angle Selection for Surfaces of Non-Metals

David Rath,[1] Vojtěch Mikerásek,[1] Chunlei Wang,[1] Moritz Eder,[1] Michael Schmid,[1] Ulrike Diebold,[1] Gareth S. Parkinson,[1] and Jiří Pavelec[1]

[1] *TU Wien, Institute of Applied Physics, Wiedner Hauptstraße 8-10/134, 1040 Vienna, Austria*

## Formulas Used for the Calculations

In the following, $r$ is the reflection coefficient for the electric field, and $R$ is the reflectivity, i.e., the ratio of the reflected vs. incoming intensity. To calculate $r$ of the clean surface for different polarizations and materials, we used the Fresnel equations stated in the work of Langreth[1] where $\epsilon = \epsilon' + i\epsilon''$ is the complex dielectric constant of the reflecting substrate and $\theta$ the incidence angle. Zero indicates the reflection on the clean surface, and p or s defines the light polarization. With $R_i = |r_i|^2$ one can calculate the reflectivity. The complex refractive index $\hat{n} = n + ik$ is related to the complex dielectric constant $\epsilon = \hat{n}^2 = (n + ik)^2$.

$$r_{0,\mathrm{p}} = \frac{\epsilon \cot(\theta) - \sqrt{-1 + \epsilon \csc(\theta)^2}}{\epsilon \cot(\theta) + \sqrt{-1 + \epsilon \csc(\theta)^2}} \tag{1}$$

$$r_{0,\mathrm{s}} = \frac{\cot(\theta) - \sqrt{-1 + \epsilon \csc(\theta)^2}}{\cot(\theta) + \sqrt{-1 + \epsilon \csc(\theta)^2}} \tag{2}$$

To consider the influence of an adsorbate on the surface, we used the following equations based on the work of Langreth.[1] $r_{\mathrm{p}\perp}$ and $r_{\mathrm{p}\parallel}$ describe the reflection coefficient for p-polarized light with the dipole moment orientation perpendicular or parallel to the surface, respectively (i.e., vertical and horizontal dipole, in the plane of incidence). The case of s-polarization and a dipole moment parallel to the surface but perpendicular to the plane of incidence is described with $r_{\mathrm{s}\parallel}$. Here, $N$ is the density of the adsorbates on the surface, $\tilde{v}$ the wavenumber and $\alpha_{i,j}$ the polarizability in volume units (see equation (6)).

$$r_{\mathrm{p},\perp} = \frac{\epsilon \cot(\theta) - \sqrt{-1 + \epsilon \csc(\theta)^2} + 8\,i\,N\,\alpha_{\mathrm{p},\perp}\,\pi^2 \epsilon\,\tilde{v} \sin(\theta)}{\epsilon \cot(\theta) + \sqrt{-1 + \epsilon \csc(\theta)^2} - 8\,i\,N\,\alpha_{\mathrm{p},\perp}\,\pi^2 \epsilon\,\tilde{v} \sin(\theta)} \tag{3}$$

$$r_{\mathrm{p},\parallel} = \frac{-i\left(-\epsilon \cot(\theta) + \sqrt{-1 + \epsilon \csc(\theta)^2}\right) + 8\,N\,\alpha_{\mathrm{p},\parallel}\,\pi^2\,\tilde{v} \cos(\theta)\sqrt{-1 + \epsilon \csc(\theta)^2}}{i\left(\epsilon \cot(\theta) + \sqrt{-1 + \epsilon \csc(\theta)^2}\right) + 8\,N\,\alpha_{\mathrm{p},\parallel}\,\pi^2\,\tilde{v} \cos(\theta)\sqrt{-1 + \epsilon \csc(\theta)^2}} \tag{4}$$

$$r_{\mathrm{s},\parallel} = \frac{\cot(\theta) - \sqrt{-1 + \epsilon\,\csc(\theta)^2} + 8\,i\,N\,\alpha_{\mathrm{s},\parallel}\,\pi^2\,\tilde{v} \csc(\theta)}{\cot(\theta) + \sqrt{-1 + \epsilon\,\csc(\theta)^2} - 8\,i\,N\,\alpha_{\mathrm{s},\parallel}\,\pi^2\,\tilde{v} \csc(\theta)} \tag{5}$$

In equation (5), two misprinted signs in equation (3.23) presented in Langreth[1] were corrected, which is in line with the calculation steps described above equation (3.23) in the manuscript. The polarizability was modeled with a Lorenzian oscillator taken from Tobin.[2] $\alpha_e$ is the electronic polarizability, $\alpha_v$ the vibrational polarizability, $\tilde{\nu}$ the wavenumber, $\tilde{\nu}_0$ the resonance frequency and $\gamma$ the line width; $\tilde{\nu}_0$ and $\gamma$ are expressed in wavenumber units.

$$\alpha_{i,j}(\tilde{\nu}) = \alpha_e + \frac{\alpha_v}{1 + \left(\frac{\tilde{\nu}}{\tilde{\nu}_0}\right)\left(\frac{\tilde{\nu}}{\tilde{\nu}_0} + i * \frac{\gamma}{\tilde{\nu}_0}\right)} \tag{6}$$

## Effects in the Real System

In real IRAS systems, several aspects change the incidence angle-dependent $\Delta R/R_0$. FIG. S1 shows $\Delta R/R_0$ when considering the effects of non-ideal polarization,[3] beam divergence,[4] and adsorbate area illumination. Comparison with the uncorrected case shows that s-polarization remains almost unaffected, which is expected due to the weak dependence on the angle. For p-polarization, the singularity at the Brewster angle gets smoothed out. $\Delta R/R_0$ shows a minor dip for all polarizations at 72° originating from the detection of intensity reflected from the adsorbate-free sample area in our setup. The degree of polarization of 99.5 % assumed in Fig. S1 is based on the data of the polarizer; it does not include polarization aberrations related to beams that are not in the plane of incidence (e.g. skew aberration).

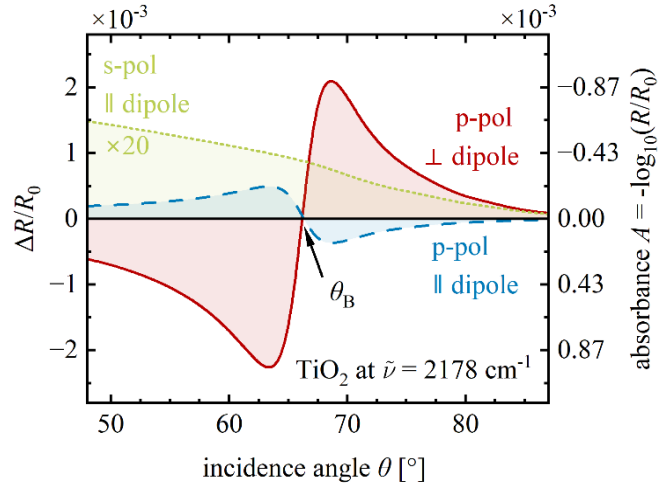

FIG. S1. Corrected calculated normalized reflectivity difference of an adsorbate on a TiO$_2$ surface. The three principal orientations of the dipole moment are shown together with the resonating orientation of the polarized light: horizontal dipole moment perpendicular to the incidence plane and s-polarization (green, dotted), horizontal dipole parallel to the incidence plane and p-polarization (blue, dashed), and vertical dipole moment and p-polarization (red). The s-polarized data are multiplied by 20. The calculation includes ±2° IR beam divergence within the incidence plane, a degree of light polarization of 99.5 %, and accounts for the light detected from the adsorbate-free area on the sample in our experimental setup. The curves are plotted for the range of incidence angles accessible in our IRAS system.

## Mirror Focal Lengths

| mirror name | type | front focal length [mm] | back focal length [mm] |
|---|---|---|---|
| spectrometer-link mirror | off-axis parabolic mirror | – | 50.8 |
| input mirror | off-axis elliptical mirror | 50.9 | 281.2 |
| illumination mirror | off-axis elliptical mirror | 409.8 | 60.2 |
| collector mirror | off-axis elliptical mirror | 50.9 | 281.2 |
| detector mirror | off-axis elliptical mirror | 308.7 | 43.4 |

Table SI. Focal lengths and mirror types of the specific mirrors used for the IRAS system. The spectrometer-link mirror is a standard commercial mirror, and all the other mirrors are custom mirrors.